**Design and User Satisfaction of Interactive Maps for Visually Impaired People,**
Anke Brock, Philippe Truillet, Bernard Oriola, Delphine Picard and Christophe Jouffrais,
In: Miesenberger, K.; Karshmer, A.; Penaz, P.; Zagler, W. (eds.) ICCHP 2012. Lecture Notes in Computer Science, Volume 7383, p. 544-551. Springer, Heidelberg (2012)

# Design and User Satisfaction of Interactive Maps for Visually Impaired People

Anke Brock[1], Philippe Truillet[1], Bernard Oriola[1],
Delphine Picard[2], Christophe Jouffrais[1]

[1]IRIT-UMR5505, Université de Toulouse & CNRS, Toulouse, France
`{brock,truillet,oriola,jouffrais}@irit.fr`
[2]Octogone, Université de Toulouse, France & Institut Universitaire de France
`delphine.picard@univ-tlse2.fr`

**Abstract.** Multimodal interactive maps are a solution for presenting spatial information to visually impaired people. In this paper, we present an interactive multimodal map prototype that is based on a tactile paper map, a multi-touch screen and audio output. We first describe the different steps for designing an interactive map: drawing and printing the tactile paper map, choice of multi-touch technology, interaction technologies and the software architecture. Then we describe the method used to assess user satisfaction. We provide data showing that an interactive map – although based on a unique, elementary, double tap interaction – has been met with a high level of user satisfaction. Interestingly, satisfaction is independent of a user's age, previous visual experience or Braille experience. This prototype will be used as a platform to design advanced interactions for spatial learning.

## 1 Introduction

Human navigation is a very complex behavior that mainly relies on vision. Indeed, vision provides the pedestrian with static and dynamic cues that are essential for position and orientation updating, estimation of distance, etc. Hence, for a visually impaired person, navigating in familiar environment is not obvious, and becomes especially complicated in unknown environments. The major problem is a lack of information concerning the environment which leads to deficits in orientation and mobility. These problems often mean that the visually impaired travel less, which influences their personal and professional life and can lead to exclusion from society. With 285 million people being visually impaired around the world [1], it is therefore a very important task to make spatial information accessible to the visually impaired.

Accessible geographic maps represent valuable assistance for journey preparation and can thus help to overcome the fear and stress related to traveling. Maps are projective two-dimensional symbolic representations of a real-space in smaller scale [2]. They can represent spaces of different dimensions (going from a room up to the whole world). Maps allow the absolute and relative localization of objects as streets or buildings, the estimation of distances and directions, as well as finding an itinerary between two points. As stated by Hatwell et al [2] this information is only accessible if the user possesses the necessary perceptual and cognitive skills allowing access to the symbolic codes of the maps. Conversely, the map must be designed so that a person with visual impairments can access the information.

Traditionally, raised-line paper maps are used to present geographic information to visually impaired people. However, Jacobson [3] mentioned that raised-line paper maps have numerous limitations. Most importantly, the map content needs to be simplified as the fingertip's resolution is less than the eye's. Furthermore, Braille is used for giving textual information, which requires a lot of space. Therefore tactile maps tend to be overloaded and thus unreadable. Besides, not every visually impaired person can read Braille. Finally, the content of such a map cannot be adapted dynamically. Multimodal interactive maps undoubtedly represent a solution to overcome these problems.

There are different concepts for interactive maps, some with auditory only, some with auditory and haptic feedback. An auditory map was proposed by Jacobson [3]. The user navigates in a model of the environment with a touchpad and receives auditory feedback. Buzzi et al proposed a similar system based on data gathered from the web [4]. Rice et al [5] added haptic feedback to an auditory map by using a force-feedback mouse. Different studies [6, 7] proposed a combination of audio and tactile output based on a matrix of refreshable pins. The 3D-Finger system [8] used image recognition to follow a user's finger during map exploration. The finger position was associated to the content of the underlying tactile paper map in order to determine the corresponding audio output. Finally, several map projects were based on a combination of touch screens and raised-line paper maps [9–12]. The user could retrieve tactile information by exploring the raised-line map. The systems gave additional audio information (e.g. street names) when the user touched the screen.

Several advantages and disadvantages exist for the different types of interactive maps. We chose to place a raised-line map on top of a touch-screen for the following reasons: First of all, most blind users are used to explore raised-line maps. The usage of the prototype is then easy to learn and relies on acquired skills. Second, Rice et al [5] proposed combining tactile and audio modalities because they may both represent spatial information but have complementary functions. For example, Braille labels can be avoided when using speech output. The map can then be designed without overcrowding, including essential tactile information only. Furthermore, when using a raised-line map, it is easy to use tactile cues (e.g. outlines of the map) for keeping mental orientation. Besides, the usage of both hands allows the user to keep fixed reference points while exploring with the other hand. On the contrary, when using a pointing device (e.g. a force-feedback mouse with a single moving cursor), it is much more difficult to keep the reference frame in mind [5]. The last argument in favor of research on an embossed paper map placed on a multi-touch table is that touch screens and raised-line printers are nowadays relatively cheap. They may actually be used by visually impaired people in associations and schools. Even more convincing is that the improvement of haptic refreshable touch screens (as for example the Surf-pad [13]) will promote the design of interactive maps, without having to superimpose raised-line paper maps.

## 2 Designing an Interactive Map for Visually Impaired People

The following paragraphs describe the different steps for designing an interactive map based on the combination of paper map, touch-screen and audio output which consist in drawing and printing the raised-line paper map, choice of multi-touch technology, interaction methods and software architecture.

### 2.1 Step 1: Drawing and Printing a Tactile Map

The design of a raised-line paper map includes two aspects: the layout of the map as well as the method for printing the map.

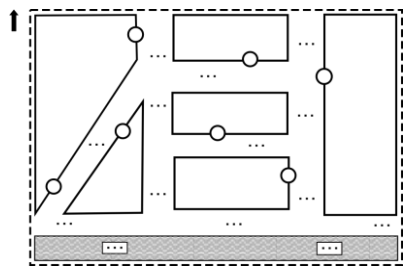

**Fig. 1.** Drawing of the tactile map used for our prototype

There are no conventions for designing tactile maps, which means that each map uses different symbols and textures. Nevertheless, there are several guidelines that rely on the specificities of the tactile modality. It is important to note that tactile resolution is inferior to visual resolution [14]. In addition, tactile perception is rather serial, whereas vision is synoptic. In contrast to audio description, tactile exploration does not follow an imposed order. Yet, as stated by Hatwell [15], this does not imply that touch is not adapted for perceiving spatial information. Tatham [14] identified a set of symbols with their minimum and maximum perceptible dimensions. Recently, Paladugu et al [16] evaluated different tactile patterns, proposing a set of symbols for tactile maps. We based our map design on these existing guidelines, and chose a set of tactile symbols that were clearly distinguishable.

For map design, we used the Inkscape editor and SVG file format. SVG is used in many different map projects (see e.g. [10]). It is based on the extensible markup language (XML) which is specified by the World Wide Web Consortium (W3C). It allows both visual and textual views, which is very convenient for adding names and description tags to geographic elements.

For this study, we designed a simple fictive map with 6 streets, 6 points of interests and one river (Fig. 1). Names of streets and points of interests were chosen according to the number of syllables and the frequency of usage as mentioned in the French-speaking database Lexique [17]. Before any experimentation, pre-tests with a blind user confirmed that the map elements were all fully accessible.

The two main methods used for printing raised-line maps are vacuum forming and microcapsule paper. Perkins [18] showed that both techniques are efficient for presenting spatial information. We chose microcapsule paper maps because it is easier to use as production material, and also because the paper used in this case is slimmer, which is advantageous to detect inputs on the touch table through the map.

### 2.2   Step 2: Selecting the Multi-Touch Screen

The multi-touch market is rapidly evolving, introducing a great number of new models and technologies. We identified requirements in order to select an adapted technology for designing an interactive map prototype (see [19] for details). Briefly, we tested different devices and we finally chose the 3M Inc. multi-touch screen (model M2256PW) relying on the projected capacitive multi-touch technology. At the time of purchase it was the only multi-touch screen functioning with a paper map placed on top of it. In addition, the size of the table was adapted for displaying various geographic maps.

### 2.3   Step 3: Interaction Technologies

The choice of input and output modalities is an important aspect of interactive map design. The relief of the tactile map was the first available sensory cue. We added the Realspeak SAPI 4.0 French text-to-speech synthesis (voice "Sophie") which possesses a good intelligibility and user appreciation [20].

Most interactive maps use simple touch events for input (see e.g. [10]). Kane et al [21] studied gesture interaction for blind people. They found that simple and double taps were easily usable. We chose a simple tap as a basic interaction method in order to validate the map layout and interactivity. Once again, we made pre-tests with two legally blind subjects who were both experienced map users. Interestingly, one of them had explored another interactive map beforehand. Although the simple tap worked fine with sighted users, it did not work with blind users. We observed that, contrary to sighted subjects, the visually impaired users explore tactile maps with several fingers. When multiple fingers were simultaneously applied on the display, many sound outputs were produced. The two blind users who tested the system were then not able to comprehend which finger caused sound outputs. This problem appeared to be specific to an interactive map based on a multi-touch surface (to our knowledge all other projects relied on mono-touch tables). We therefore implemented a double-tap as input interaction, which proved to be efficient to interact with the prototype.

### 2.4   Step 4: Software Architecture

There exist multiple application programming interfaces for multi-touch devices. As we needed to directly access the touch events, we used the touch-screen low level driver. For each touch event, we obtained an ID (automatically reused when free), the (x; y) coordinates and a timestamp. These data were used for online interaction and logged in a data file for offline analysis. The software architecture of the prototype was made of different software modules connected via the Ivy middleware [22] (see [19] for details). This architecture is very versatile as it allows replacing software modules. The prototype can then easily be adapted with different hardware, maps or experimental requirements.

## 3 Testing User Satisfaction for the Interactive Map

### 3.1 Experimental Protocol

In our study, we assessed the user satisfaction concerning the interactive map with the SUS questionnaire [23] translated into French. As proposed by Bangor et al [24] we replaced the usage of the word "cumbersome" by "awkward" to make question 8 of the SUS easier to understand. In an earlier study we had observed negative reactions to the question 7 "I would imagine that most people would learn to use this product very quickly". Users remarked that "most people" would not use a product for visually impaired people. Therefore, we proposed "I think that most visually impaired people would learn to use this product very quickly".

Twelve legally blind users (6 men, 6 women) were involved in the experiment. All users possessed prior experience with regular tactile paper maps and were Braille readers. Each user attended an individual session with one experimenter. The session started with a familiarization phase during which the user explored a map similar to the one used for testing. The experimenter checked that the user was used to the double-tap interaction technique. Next, the experimenter interviewed the user on personal characteristics (chronological age, Braille experience and age at onset of blindness; see Table 1). Then, the instruction was to explore and learn the interactive map as quickly and accurately as possible. The user finally completed the SUS questionnaire and was asked to describe the aspects that he particularly enjoyed or disliked during exploration.

### 3.2 Results

SUS scores were calculated according to Brooke [23]. Results (see Table 1) provided evidence for a high user satisfaction concerning the interactive map. The mean value of the scores was 87.3 (SD = 15.1). Bangor et al [24] considered scores above 85 as "excellent". The maximum score obtained was 97.5. All scores were superior to 75 points (thus at least "good" [24]) with exception for one subject whose score was 45. Users' characteristics varied significantly according to age (from 21 to 64 years), age at onset of blindness (congenitally blind, late-blind including a user who only lost sight some years ago), and Braille reading experience (from 5 to 58 years). However, we did not observe any correlation between the SUS scores and at least one of these characteristics.

Most users quickly learned the double-tap, whereas the user who gave a SUS score of 45 encountered problems using the double-tap. This user (female, aged 64) possessed prior experience with paper maps with Braille legends and almost 60 years of experience in Braille reading. She mentioned that she enjoys reading Braille and that she had been 'surprised' with the usage of an interactive map.

We asked users for the aspect that they most enjoyed or disliked about the interactive map prototype. As positive aspects they stated that it did not require reading Braille or that they generally prefer speech output (3 users), that there was no need to read a legend (1 user), that it was easy to memorize (1 user), that it was easy to use (1 user) and that usage was ludic (1 user). Aspects they did not like concerned interaction problems with the map (1 user) or that they could more easily memorize written information (1 user).

In addition to this user group, we met a very interesting case: a participant aged 84 years who lost sight at the age of 66. Hence, he learnt Braille lately and had limited reading skills. A standard raised-line map with Braille text was not accessible for him, unless we printed the Braille with large spacing between letters. Contrary, he could immediately use the interactive map and gave an excellent score of 87.5 points in the SUS questionnaire. The interactive map provided him with access to spatial information that he could not have obtained with a regular paper map.

Table 1. Personal characteristics and SUS scores for each user

| User | 1 | 2 | 3 | 4 | 5 | 6 | 7 | 8 | 9 | 10 | 11 | 12 |
|---|---|---|---|---|---|---|---|---|---|---|---|---|
| Gender | F | F | M | M | F | M | M | F | F | F | M | M |
| Age | 31 | 58 | 25 | 21 | 33 | 53 | 31 | 54 | 38 | 64 | 48 | 59 |
| Onset of blindness (age) | 2 | 15 | 0 | 14 | 26 | 19 | 5 | 0 | 0 | 10 | 25 | 0 |
| Braille experience (years) | 25 | 46 | 19 | 6 | 5 | 35 | 25 | 48 | 26 | 58 | 22 | 49 |
| SUS score | 90 | 97,5 | 95 | 95 | 80 | 90 | 97,5 | 95 | 97,5 | 45 | 75 | 90 |

Note: Age at onset of blindness corresponds to the age of legal blindness and not to the first occurrence of visual impairment.

## 4 Conclusion and Future Work

In this article we presented the design and we evaluated the satisfaction related to an interactive map prototype based on a double-tap interaction. It appears that the prototype is very versatile and is an ideal platform to design more advanced interactions. In addition, despite important inter-individual differences, SUS scores provided evidence for a general high satisfaction. First we may note that the interactive map is satisfactory independently of chronological age (range from 21 to 64 years old), which is counterintuitive as one might think that the older are more refractory to technologies. Interestingly, the satisfaction was also excellent independently of the age at onset of blindness. This is important as we know that the age at onset of blindness has important outcomes in terms of adaptation to blindness (general operation, mental imagery, etc.). Finally, the satisfaction was also excellent regardless of Braille experience, except for one user. This result shows that even experts that are particularly attached to Braille are not reluctant to a sound based technology. In addition, the interactive map provides poor Braille readers with access to spatial information.

Satisfaction is one component of usability. Efficiency and efficacy of the interactive map also have to be evaluated. Relying on this prototype, we are designing an experiment to measure satisfaction, efficiency (exploration time) and efficacy (spatial learning) of the interactive map. We aim to show that all three components of usability are higher for interactive maps than for regular (paper) maps with Braille legends.

A second aspect of our work currently consists in designing advanced interactions. Our observations showed that blind users perform specific haptic exploration strategies that impose adapted interaction. Interaction must, especially, be distinguished from regular map exploration and should promote spatial learning. Currently, most interactive maps use mono-touch displays that present important limitations concerning interaction (i.e. simple tap events only). Multi-touch displays would enable new possibilities based on multiple fingers or gestural interaction.